\documentstyle[preprint,aps,12pt]{revtex}

\begin{document}

\title{Stochastic resonance in bistable systems: The effect of 
simultaneous additive and multiplicative correlated noises}

\author{Claudio J. Tessone \thanks{Electronic address: 
tessonec@cab.cnea.edu.ar, Fellow of CNEA} 
and Horacio S. Wio\thanks{Electronic address: wio@cab.cnea.edu.ar, 
Member of CONICET, Argentina and Regular Associate ICTP. 
http://www.cab.cnea.edu.ar/Cab/invbasica/FisEstad/estadis.htm }}
\address{Comisi\'on Nacional de Energ\'{\i}a At\'omica,\\
Centro At\'omico Bariloche and Instituto Balseiro (CNEA and 
UNC),\\
8400-San Carlos de Bariloche, Argentina}

\def\ave#1{\langle #1 \rangle}

\maketitle

\begin{abstract}
We analyze the effect of the simultaneous presence of correlated 
additive and multiplicative noises on the stochastic resonance response 
of a modulated bistable system. We find that when the correlation 
parameter is also modulated, the system's response, measured through the 
output signal-to-noise ratio, becomes largely independent of the additive 
noise intensity. 
\end{abstract}

\newpage

The phenomenon of {\it stochastic resonance} (SR) has attracted considerable 
interest in the last decade due, among other aspects, to its potential 
technological applications for optimizing the output signal-to-noise ratio 
(SNR) in nonlinear dynamical systems. The phenomenon shows the 
counterintuitive role played by noise in nonlinear systems as it 
contributes to enhance the response of a system subject to a weak external 
signal. There is a wealth of papers, conference proceedings and reviews on 
this subject, Ref. \cite{gamma1} being the most recent one, showing the 
large number of applications in science 
and technology, ranging from paleoclimatology, to electronic circuits, 
lasers, and noise-induced information flow in sensory neurons in living 
systems. 

A tendency shown in recent papers, and determined by the possible 
technological applications, points towards achieving an enhancement 
of the system response (that is obtaining a larger output SNR) by means of 
the coupling of several SR units \cite{buls1,jung,buls2,wio,fabio} in what 
conforms an ``extended medium" \cite{wiocas}. Another aspect that has also 
attracted interest is finding a system where the SNR becomes largely 
independent of external parameters such as noise intensity 
\cite{buls2,coll}.

In this work we present some results related to the latter 
aspect. In order to reach such a goal we study a bistable system subject to both
an additive and a multiplicative noise but, at variance with the work in 
Ref. \cite{gamma2}, we consider that both noises are correlated. In addition 
to the modulation of the bistable potential by a weak external signal, we 
also consider the effect of a modulation of the correlation between both 
noises. We show that the result of this contribution is to widen the maximum 
of the SNR as a function of the additive noise intensity, making the detection 
of the signal less sensitive to the actual value of that noise. It is worth 
remarking here that on one hand, the additive (external) noise can be 
(naturally) assumed as white, while on the other hand, the multiplicative 
(internal) one, that generally involves characteristic time scales of 
the system, would not necessarilly be white. However, as indicated in 
\cite{mdrr}, as a first step in this study, we can approximate the 
multiplicative colored noise by a white one.
Here we exploit the results of Madureira {\em et al.} \cite{mdrr} where it 
was shown that the activation rate in a bistable system subject to correlated 
additive and multiplicative noises is dramatically suppressed or enhanced 
according to the sign of the correlation. 
The model equation we consider corresponds to a onedimensional bistable system 
described by the Langevin equation
\begin{equation}
\dot{x} = \varepsilon (t) + x - x^3 + x \xi_1(t) + \xi_2(t),
\end{equation}
where $\varepsilon (t) = \varepsilon _0 \cos(\Omega t)$, $\varepsilon _0$ and $\Omega$ 
are the intensity and frequency of the potential modulation respectively. The additive 
and multiplicative Gaussian white noises, indicated by $\xi_1(t)$ and $\xi_2(t)$, satisfy 
\begin{eqnarray}
\ave{\xi_i(t)} &=& 0, \hspace{2cm} i=1,2 \nonumber \\
\ave{\xi_1(t) \xi_1(t')} &=& 2 \delta(t-t') Q \nonumber \\
\ave{\xi_2(t) \xi_2(t')} &=& 2 \delta(t-t') D \nonumber \\
\ave{\xi_1(t) \xi_2(t')} &=& 2 \rho (t)\, \delta(t-t') \sqrt{D\,Q}. 
\end{eqnarray}
The correlation intensity between both noises is indicated by the parameter 
$\rho (t)$, that fulfills the condition $|\rho| \leq 1$. 
The associated the Fokker-Planck equation (Stratonovich prescription)
\begin{eqnarray}
\partial_{t} P(x,t) = &-& \partial_x
\left[ \left( x - x^3 + Q x + \rho(t) \sqrt{Q\,D} - \varepsilon(t) \right) 
P(x,t) \right] \nonumber \\ 
&+& D \, \partial^2_{xx} \left[ \left( 1 + R x^2 + 2 \rho(t) \sqrt{R} \right)
P(x,t) \right].  \label{fpe1part} 
\end{eqnarray}
In what follows we assume that $\rho$ is a time dependent function 
having the form $\rho (t)= \rho_0 \, \cos(\Omega_{\rho} t)$. Here, and in order 
to simplify the analysis, we consider the case when $\Omega_{\rho} = \Omega$, 
the most general case will be discussed elsewhere \cite{tess}.

In order to perfom the evaluation of the correlation function and power 
spectral density needed to obtain the SNR, we exploit the results of the 
two-state model  \cite{nico,mcnmr}. Those authors have 
reduced the problem of obtaining the SNR of a nonlinear and essentially 
bistable system subject to a weak periodic signal, to a description where 
the transitions occur between the two minima of the deterministic 
potential. The main approximation is an adiabatic-like one, corresponding  
to the assumption that the relaxation times around each minima are much 
shorter than any other characteristic time, such as the Kramers time for 
transitions between the two stable points or the inverse of the signal 
frequency $\Omega$. It is also important to note here that, if the modulation 
is done around $\rho \neq 0$, it becomes necessary to extent the two state 
approach in order to include the asymmetry of the potential. 

In the absence of any signal, the deterministic potential of the system has two 
minima located at the points $x_{\pm}= \pm 1$. It has been shown that such 
minima are not coincident with the maxima of the stationary probability 
distribution \cite{flnsk}. 

Let us call $W_{+}(t)$ and $W_{-}(t)$ the nonstationary 
transition rates from the state $x_{+}$ to $x_{-}$ and from the state 
$x_{-}$ to $x_{+}$ respectively. 
As indicated before, the signal is introduced through a simultaneous 
modulation in the potential and in the correlation. We may note here that 
the last contribution does not modify the deterministic potential barrier.  
However, what we find is that the rates $W_{\pm}$ do change, i.e. 
in \cite{mdrr} it was shown that if $\rho > 0$ the rate $W_{+}$ increases
while $W_{-}$ decreases, and viceversa.  

If both modulations are small in comparison with the barrier height, that is 
$\rho_0 \ll 1$ and $ \varepsilon_0 \ll V(0) - V(\pm 1)$, 
then it is possible to make a Taylor expansion of the functions 
$W_{\pm}(t)$. We thus obtain 
\begin{equation} 
\label{frecuencia}
W_{\pm}(t) = \frac{1}{2} \left(W_{o} \mp (\alpha_{\rho} + \alpha_{\varepsilon})
 \cos (\Omega t) + {\cal O}(\varepsilon_0 ^2) + {\cal O}(\rho_0^2) \cdots \right), 
\end{equation}
where $W_{o}$ is the transition rate evaluated in absence of modulation, 
$\rho_0=\varepsilon_0=0$. The terms $\alpha_{\rho}$ 
and $\alpha_{\varepsilon}$ are given by 
\begin{eqnarray}
\frac{\alpha_{\varepsilon}}{2} &=& - \left.\frac{d W_o}{d \varepsilon}\right| _{\varepsilon=0} \, \varepsilon_0 \label{alfa1} \\
\frac{\alpha_{\varepsilon}}{2} &=& - \left.\frac{d W_o}{d \rho}\right| _{\rho=0} \, \rho_0 \label{alfa}. 
\end{eqnarray}
The transition rate $W_o$ could be obtained calculating the 
mean-first-passage-time $T(R,\rho)$ \cite{mdrr}, yielding 
\begin{eqnarray}
\label{mfpt}
\frac{1}{W_o} = T(R,\rho) = &\frac{1}{D}& \int_{-1}^1 \, dx \, H(x) \, \exp\left[ \Phi(x) / D \right] \nonumber \\
& \,\, & \int_{-\infty}^{x} \, dy \, H(y) \, \exp\left[ - \Phi(y)/D \right].
\end{eqnarray}
where $R \equiv Q/D$, and the function $H(x)$ is
$$
H(x) = \sqrt{\frac{1}{1+R x^2+2\rho \sqrt{R} x}}.
$$
The determination of the effective potential $\Phi(x)$ is made studying 
the stationary probability density, which is given by 
$$
\Phi(x) = \int_{-\infty}^{x} H(x')^2 \, \left( 1+R x'^2 + \left( \varepsilon_0 \cos (\Omega t) + 2 \rho \sqrt{R} \right) x' \right) dx'.
$$

We obtained the derivatives $dW_o(R,\rho)/d\varepsilon$ and 
$dW_o(R,\rho)/d\rho$ from the evaluation of $dT(R,\rho)/d\varepsilon$ and 
$dT(R,\rho)/d\rho$ respectively, through
\begin{eqnarray}
\frac{dW_o(R,\rho)}{d\varepsilon} = - \frac{1}{T(R,\rho)^2} \frac{dT(R,\rho)}{d\varepsilon} \label{omega1}\\
\frac{dW_o(R,\rho)}{d\rho} = - \frac{1}{T(R,\rho)^2} \frac{dT(R,\rho)}{d\rho} \label{omega2}. 
\end{eqnarray}
The results for the derivatives of $T(R,\rho)$, after a long but 
straightforward calculation, are  
\begin{eqnarray}
D^2 \frac{d T}{d \varepsilon} = 
 &- & \int_{-1}^{1} \, dx \, H(x) \exp\left[\Phi(x)/D \right] 
\int_{-1}^{x} \, dy \, H(y) \, \exp\left[ - \Phi(y)/D \right]
\int_{-1}^{x} \, dy \, H(y)^2 \, y \nonumber \\
& + & \int_{-1}^{x} \, dy \, H(y) \exp\left[-\Phi(y)/D \right]
 \int_{-1}^y  \, dz \,  H(z)^2 \, z \label{domdvare}, 
\end{eqnarray}
for the contribution of the potential modulation, while for the contribution 
of the correlation modulation we obtain 
\begin{eqnarray}
\frac{D^2}{\sqrt{QD}} \frac{d T}{d \rho} = & - & \int_{-1}^{1} \, dx \, H(x) 
\exp\left[\Phi(x)/D \right] \times H(x)^2 \,x \, \int_{-1}^{x} \, dy \, H(y) \, 
\exp\left[ - \Phi(y)/D \right] \nonumber \\
& + & \frac{2}{D} \int_{-1}^{x} \, dy \, H(y) \, 
\exp\left[ - \Phi(y)/D \right] \int_{-1}^{x} \, dy \, y^2 (1-y^2) H(y)^4 \nonumber \\
& - & \int_{-1}^{x} \, dy \, H(y)^3 y + 
 \frac{2}{D} \int_{-1}^{x} \, dy \, H(y) 
\exp\left[-\Phi(y)/D \right] \int^{y}_{-1} z^2 (1-z^2) H(z)^4 . \label{domdrho}
\end{eqnarray}

Once $\alpha_{\varepsilon}$ and $\alpha_{\rho}$ are determined, we can obtain 
the auto-correlation function, which is given by 
\begin{eqnarray}
\ave{x(t)|x(t+\tau)} = & \, & e^{- W_o |\tau|} \left[ 1- \frac{(\alpha_{\varepsilon} + \alpha_{\rho})^2 \cos(\Omega t - \phi)}{\alpha + \Omega} \right] \nonumber \\
&+& \frac{(\alpha_{\varepsilon} + \alpha_{\rho})^2 \left( 
\cos (\Omega \tau) + \cos \left[ \Omega (2t+\tau) +2\phi \right]) 
\right) }{2 \left( (\alpha_{\varepsilon} + \alpha_{\rho})^2 + \Omega^2\right)}. 
\end{eqnarray}
From $\ave{x(t)|x(t+\tau)}$ the power-spectrum density (psd) 
$\ave{S(\omega)}$ is readily obtained as 
\begin{eqnarray}
\ave{S(\omega)}_t = 
\left[ 1-\frac{(\alpha_{\rho}+
\alpha_{\varepsilon})^2}{2(W_o^2+\Omega^2)} \right] 
\left[ \frac{2W_o}{W_o^2+\omega^2} \right] + 
\frac{\pi (\alpha_{\rho}+
\alpha_{\varepsilon})^2}{2(W_o^2+\omega^2)} \delta(\omega-\Omega) 
\end{eqnarray}
where we have written only the expression for positive values of $\omega$ 
\cite{mcnmr}.

Then, to determine the output SNR ${\cal R}$, we use the standard definition 
\begin{equation}
\label{snr}
{\cal R} =10 \, \log _{10}\left(\frac{S_s }{S_n \Delta} + 1\right), 
\end{equation}
where $S_s$ and $S_n$ are the (integrated) psd with and without signal respectively, 
evaluated at the modulation frequency. 
Here, the parameter $\Delta$ is introduced in order to tune the theoretical 
result when compared with a numerical simulation or an experiment, $\Delta$ 
being related to the bandwidth of the sampling frequencies.
It is clear that the evaluation of ${\cal R}$ is critically dependent 
on the evaluation of $dW_o / d\varepsilon$ and $dW_o / d\rho$. For this 
reason the results of Eqs. (\ref{domdvare}, \ref{domdrho}) have been 
numerically tested for a large range of values of the different parameters.

One of the goals of the present work was to look for a larger independence 
of the SNR from the external parameters, particularly the additive noise
intensity. In Fig. 1 we show the results for ${\cal R}$ as a 
function of the noise intensities $D$ and $Q$, for a fixed value of 
$\rho _0$ and different values of $\varepsilon _0$. The widening of the 
peak as a function of $D$ for a region of values of $Q$ is apparent. 
However, this widening occurs at the expense of a (small) reduction in the 
maximum of ${\cal R}$. The limiting case for $Q \to 0$, reducing to 
the usual SR behaviour \cite{mcnmr}, should be compared with the case when 
$Q \neq 0$ to see clearly the widening effect. Also, the behaviour for 
$D \to 0$ (and $Q \neq 0$) reduces to the case studied in \cite{gamma2}. 
Such a widening of ${\cal R}$ is due to the additive dependence of the ouput 
signal on $\rho_0^2$. 

Another novel aspect of our results is that when $\varepsilon _0 \to 0$, 
we have a SR phenomenon purely due to the modulation of the correlation 
parameter that is more localized in the $(D,Q)$ parameter space. When 
one or both of these parameters goes to $0$ or $\infty$ we find that 
${\cal R} \to 0$. 

The results obtained by modifying only the intensity of correlation modulation 
amplitude are depicted in Fig. 2. In this case, the 
results are essentially the same as in the previous figure.

It is worth to note here that in Figs. 1b,c and 2b, keeping constant $D$, the SNR 
grows as $Q$ increases. This remarkable fact implies that the additive SR (the case 
$Q=0$) could be enhaced by adding a multiplicative noise.

We have also studied the effect of changing the frequency modulation 
$\Omega$ on SNR, observing the same dependence on this parameter, as in 
other SR sytems. For very low frequencies the effect is slightly larger 
than for higher frequencies, but it is essentially frequency-independent.

Finally, in Fig. 3, we depict a contour plot of the SNR 
as a function of  $\rho_0$ and $\varepsilon_0$. We fixed $D, \, Q$ and 
the frequency modulation $\Omega$. A paraboloid-like dependence
on those variables is obtained, with ${\cal R}$ increasing as any of the 
variables increases too.

In conclusion, we have shown that the simultaneous modulation of the 
potential and the correlation parameter between the additive and 
multiplicative noises induces a widening of the output SNR, making the 
system response less dependent on the precise value of the additive 
noise. It is worth noting that if the modulation frequencies for
the potential and the correlation are too different, what we find
is a superposition of the SR effect associated to each separated modulation,
that is: two differenciated peaks in the SNR at the corresponding 
frequencies.

The present result could be of relevance for both technological and 
biological systems, as in electronic signal detectors or sensory systems 
one wishes that the detection capacity shall be as little dependent as 
possible on the (usually external) additive noise, the (usually 
internal) multiplicative noise being a kind of tuning parameter. A more complete 
study of the dependence of this effect on the modulation frequency of the 
correlation will be presented elsewhere \cite{tess}. \\
{\bf Acknowledgments:} The authors thank P. H\"anggi and  G. Nicolis for useful 
comments and V. Grunfeld for a revision 
of the manuscript. Partial support from CONICET, Argentina, through grant 
PIP Nro.4593/96, ANPCyT, Argentina, through grant 03-00000-00988, and CEB, 
Bariloche, Argentina, are also acknowledged.

\newpage

\begin{figure}[tbp]
\label{3d1}
\caption{Here we depict the SNR as function of the additive and 
multiplicative noise intensities $D$ and $Q$. On the left side we 
show the 3D plots associated with the contour plots on the right 
side. We have fixed $\rho _0 = 0.05$ and the frequency modulation 
$\Omega = 0.005$, while the different cases correspond to: 
(a) $\varepsilon _0 = 0.01$, (b) = 0.05, 
(c) = 0.10. For the sake of clarity, the point of view in the left 
side 3D plots, is from the upper left corner in the contour plots. }
\end{figure}

\begin{figure}[tbp]
\label{3d2}
\caption{We show the SNR in $(D, Q)$ 
space. Once again, the left side 3D plots correspond to the contour 
plots on the right side, and the point of view for the 3D plots corresponds 
to the upper left corner in the contour plots. We fixed the frequency 
modulation $\Omega=0.005$, the amplitude of potential modulation 
$\varepsilon_0=0.05$, and varied the potential amplitude: 
(a) corresponds to $\rho_0 = 0.01$; (b) = 0.10.} 
\end{figure}

\begin{figure}[tbp]
\label{intensidad}
\caption{Contour of ${\cal R}$ as a function of the modulation 
intensities $\rho_0$ and $\varepsilon_0$. The values for the noises are 
$D=0.25, \, Q=0.10$ and for the frequency modulation $\Omega=0.005$.}
\end{figure}

\end{document}